\documentclass{article}

\usepackage{arxiv}
\usepackage[section]{placeins}
\usepackage[utf8]{inputenc} 
\usepackage[T1]{fontenc}    
\usepackage{hyperref}       
\usepackage{url}            
\usepackage{booktabs}       
\usepackage{amsfonts}       
\usepackage{nicefrac}       
\usepackage{microtype}      
\usepackage{lipsum}		
\usepackage{graphicx}
\usepackage{xcolor}
\usepackage{doi}
\usepackage{algorithm}
\usepackage{algpseudocode}
\usepackage{amsmath}
\usepackage{xcolor} 

\usepackage[backend=biber,style=numeric,citestyle=chem-acs]{biblatex}
\usepackage{amssymb}
\DeclareFontFamily{U}{stix2bb}{}
\DeclareFontShape{U}{stix2bb}{m}{n} {<-> stix2-mathbb}{}

\NewDocumentCommand{\indicator}{}{\text{\usefont{U}{stix2bb}{m}{n}1}}

\title{Active and transfer learning with partially Bayesian neural networks for materials and chemicals}

\author{ \href{https://orcid.org/0000-0002-1101-3160}{\includegraphics[scale=0.06]{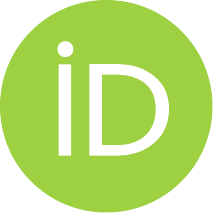}\hspace{1mm}Sarah I. Allec} \\
        Physical Sciences Division\\
	Pacific Northwest National Laboratory\\
	Richland, WA 99354 \\
	\texttt{sarah.allec@pnnl.gov} \\
	\And
	\href{https://orcid.org/0000-0003-2570-4592}{\includegraphics[scale=0.06]{orcid.pdf}\hspace{1mm}Maxim Ziatdinov} \thanks{Corresponding author} \\
	Physical Sciences Division\\
	Pacific Northwest National Laboratory\\
	Richland, WA 99354 \\
	\texttt{maxim.ziatdinov@pnnl.gov} \\
}

\providecommand{\headeright}{}
\providecommand{\undertitle}{}
\providecommand{\shorttitle}{}
\renewcommand{\shorttitle}{\emph{Active and transfer learning with Bayesian neural networks for materials and chemicals}}

\hypersetup{
pdftitle={Active and transfer learning with Bayesian neural networks for materials and chemicals},
pdfsubject={cs.LG, cond-mat.mtrl-sci},
pdfauthor={Sarah I. Allec, Maxim Ziatdinov},
pdfkeywords={Active learning, Bayesian neural networks, uncertainty quantification, materials informatics, cheminformatics},
}

\begin{document}
\maketitle

\begin{abstract}
Active learning, an iterative process of selecting the most informative data points for exploration, is crucial for efficient characterization of materials and chemicals property space. Neural networks excel at predicting these properties but lack the uncertainty quantification needed for active learning-driven exploration. Fully Bayesian neural networks, in which weights are treated as probability distributions inferred via advanced Markov Chain Monte Carlo methods, offer robust uncertainty quantification but at high computational cost. Here, we show that partially Bayesian neural networks (PBNNs), where only selected layers have probabilistic weights while others remain deterministic, can achieve accuracy and uncertainty estimates on active learning tasks comparable to fully Bayesian networks at lower computational cost. Furthermore, by initializing prior distributions with weights pre-trained on theoretical calculations, we demonstrate that PBNNs can effectively leverage computational predictions to accelerate active learning of experimental data. We validate these approaches on both molecular property prediction and materials science tasks, establishing PBNNs as a practical tool for active learning with limited, complex datasets.
\end{abstract}


\section{Introduction}
Active learning (AL) \cite{Cohn_AL, Settles_AL} optimizes exploration of large parameter spaces by strategically selecting which experiments or simulations to conduct, reducing resource consumption and potentially accelerating scientific discovery \cite{Cao_AL_alloys, Lookman_AL, Wang_AL, Xu_AL, Slautin_AL, Ziatdinov_AL}. A key component of this approach is a surrogate machine learning (ML) model, which approximates an unknown functional relationship between structure or process parameters and target properties. At each step, the model uses the information gathered from previous measurements to update its 'understanding' of these relationships and identify the next combinations of parameters likely to yield valuable information. The success of this approach critically depends on reliable uncertainty quantification (UQ) in the underlying ML models.

The development of effective ML models for active learning builds upon broader advances in machine learning across materials and chemical sciences, tackling problems including phase stability \cite{Arróyave_ML_phase_stab, Peivaste_ML_phase_stab, Liu_ML_phase_stab}, thermal conductivity \cite{Huang_ML_thermal, Luo_ML_thermal, Barua_ML_thermal, Carrete_ML_thermal}, glass transition temperatures \cite{Liu_ML_Tg, Zhang_ML_Tg, Armeli_ML_Tg, Galeazzo_ML_Tg, Uddin_ML_Tg}, dielectric properties \cite{Hu_ML_dielec, Dong_ML_dielec, Grumet_ML_dielec, Shimano_ML_dielec}, and more \cite{Morgan_ML_other, Chong_ML_other, Zhong_ML_other, Schmidt_ML_other}.
However, traditional ML models often lack inherent robust UQ, requiring additional post-hoc UQ methods such as the computation of jackknife variances for random forest \cite{RF_Jackknife} or temperature scaling for neural networks \cite{NNcalibr}. These challenges often limit their application in AL workflows.
Moreover, many of them are trained on computational data, such as density functional theory calculations \cite{FAIRification,Data_Mngmt_Challenges,TL_Comp_to_Expt},
and generalization to experimental workflows in physical labs, where data are often sparse, noisy, and costly to acquire, is often non-trivial and requires predictions with reliable coverage probabilities.

Gaussian Process (GP) \cite{Rasmussen_GPs, Snoek_GPs, Gramacy_GPs} is an ML approach that provides mathematically-grounded UQ and has become a popular choice for scientific applications, including AL frameworks \cite{deringer2021gaussian, Ziatdinov_AL}. However, GPs struggle with high-dimensional data, discontinuities, and non-stationarities, which are common in physical science problems. Deep kernel learning (DKL) \cite{Calandra_DKL, Wilson_DKL_a, Wilson_DKL_b} attempts addressing these issues by combining neural network representation learning with GP-based UQ. While DKL has shown promise in chemistry and materials science \cite{Singh_DKL_Rxn_Outcomes, Duhrkop_DKL_FPs, Valleti_DKL_Ferro}, it is still limited by GP scalability in feature space, potential mode collapse, and conflicting optimization dynamics between its GP and neural network components \cite{Ober_DKL_pitfalls}. These limitations highlight the need for further advancement of methods to support AL in non-trivial materials design and discovery tasks.

Bayesian neural networks (BNNs), where all network weights are treated as probability distributions rather than scalar values \cite{Titterington_BNNs, Lampinen_BNNs}, offer a promising approach that combines powerful representation learning capabilities with reliable UQ. By maintaining a distribution over network parameters rather than point estimates, BNNs naturally account for model uncertainty, and are particularly effective for smaller and noisier datasets. However, reliable Bayesian inference requires computationally intensive sampling methods, making fully Bayesian neural networks prohibitively expensive for many practical applications.

In this work, we explore \textit{partially} Bayesian neural networks (PBNNs) for active learning of molecular and materials properties. We show that by making strategic choices about which layers are treated probabilistically we can achieve performance on active learning tasks comparable to fully Bayesian neural networks at significantly reduced computational cost. Furthermore, we demonstrate how PBNNs can be enhanced through transfer learning by initializing their prior distributions from weights pre-trained on computational data. We validate these approaches on several benchmark datasets, demonstrating the practical potential of PBNNs for materials and molecular design with limited, complex data.

\section{Methods}
\label{sec:methods}

\subsection{Bayesian neural networks}
In conventional, non-Bayesian NNs, network weights $\theta$ are optimized to minimize a specified loss function, resulting in a deterministic, single-point prediction for each new input. Due to their architectural flexibility they can be powerful function approximators, but are known to suffer from overfitting on small or noisy datasets and overconfidence on out-of-distribution inputs \cite{Nguyen_OOD_NNs, Hendrycks_OOD_NNs, Lakshminarayanan_OOD_NNs}. In contrast, in BNNs the weights $\theta$ are treated as random variables with a prior distribution $p(\theta)$. This not only helps reduce overfitting, but also provides robust prediction uncertainties. Given a dataset $\mathcal{D}=\{x_{i}, y_{i}\}_{i=1}^n$, a BNN is defined by its probabilistic model:
\begin{align}
& \text{\textit{Weights}: } \theta \sim p(\theta) \quad \text{ (typically } \mathcal{N}(0,1)\text{)} \label{eq:bnn_weights} \\[6pt]
& \text{\textit{Noise}: } \sigma \sim p(\sigma) \quad \text{ (typically } \text{Half-Normal}(0,1)\text{)} \label{eq:bnn_noise} \\[6pt]
& \text{\textit{Likelihood}: } y_i|x_i,\theta,\sigma \sim \mathcal{N}(g(x_i;\theta),\sigma^2) \label{eq:bnn_likelihood}
\end{align}
where $g(x_i;\theta)$ represents the neural network function mapping inputs to outputs using weights $\theta$. While we focus on normal likelihoods here for regression tasks, the framework naturally extends to other distributions (e.g., Bernoulli for classification, Poisson for count data) depending on the problem domain. The posterior predictive distribution for new input $x^*$ is then given by
\begin{equation}
p(y|x^*,\mathcal{D}) = \int_{\theta,\sigma} p(y|x^*,\theta,\sigma)p(\theta,\sigma|\mathcal{D})d\theta d\sigma \label{eq:bnn_predictive}
\end{equation}
This predictive distribution can be interpreted as an infinite ensemble of networks, with each network's contribution to the overall prediction weighted by the posterior probability of its weights given the training data. Unfortunately, the posterior $p(\theta,\sigma|\mathcal{D})$ in Eq.~(\ref{eq:bnn_predictive}) is typically intractable. It is therefore common to use Markov Chain Monte Carlo (MCMC) \cite{Hastings_MCMC} or variational inference \cite{VI2017review} techniques to approximate the posterior. The advanced MCMC methods, such as Hamiltonian Monte Carlo (HMC) \cite{hmcintro2018}, generally provide higher accuracy than variational methods for complex posterior distributions \cite{Gelman_VIvsMCMC}. Here, we employ the No-U-Turn Sampler (NUTS) extension of the HMC, which efficiently explores the posterior distribution $p(\theta, \sigma|\mathcal{D})$ of neural network parameters, especially in high-dimensional spaces, without requiring significant manual tuning \cite{Homan_NUTS}. The predictive mean ($\mu_{post}$) and predictive variance ($U_{post}$) at new data points are then given by:
\begin{align}
\mu^{post} &= \frac{1}{N} \sum_{i=1}^N g(x^*; \theta_i) \label{eq:predictive_mean} \\[6pt]
U^{post} &= \frac{1}{N} \sum_{i=1}^N (y^*_i - \mu^{post})^2 
\label{eq:predictive_variance} \\[6pt]
y^*_{i} &\sim \mathcal{N}(g(x^*; \theta_i), \sigma^2_i) \label{eq:predictive_sample}
\end{align}
where $y^*_{i}$ is a single sample from the model posterior at new input \( x^* \), $\{\theta_i, \sigma_i\}_{i=1}^N$ are samples from the MCMC chain approximating $p(\theta, \sigma|\mathcal{D})$, and $N$ is the total number of MCMC samples. Note that $U^{post}$ naturally combines both epistemic uncertainty (from the variation in network predictions across different weight samples $\theta_i$) and aleatoric uncertainty (from the noise terms $\sigma_i$), providing a comprehensive measure of predictive uncertainty \cite{Uncertainties_BDL}.

\subsubsection{Partially Bayesian neural networks}
Even with sampling methods, full BNNs can be computationally expensive for reasonably-sized datasets, in terms of number of samples or feature dimensions \cite{BNN_Intractability_1, BNN_Intractability_2, BNN_Intractability_3, BNN_Intractability_4}. Variational inference, a common approximation method for BNNs, aims to alleviate these costs but often struggles with limited expressivity \cite{VIBNNexpressiveness}, underestimation of uncertainty \cite{VI_vs_MCMC}, and sensitivity to initialization and hyperparameters \cite{VIBNNinitialization}, which degrades its performance on real-world tasks. To leverage the representational power and computational efficiency of deterministic NNs \textit{and} the advantages of BNNs, we explore partially Bayesian neural networks (PBNNs), where only a selected number of layers are probabilistic and all other layers are deterministic. Building upon existing research that proposed usage of selectively stochastic layers \cite{sharma2023bayesianneuralnetworksneed, harrison2024variationalbayesianlayers}, our work specifically investigates the potential of PBNNs in active and transfer learning contexts, with a focus on molecular and materials science datasets.

\begin{figure}[h]
    \centering
    \includegraphics[width=1.0\textwidth]{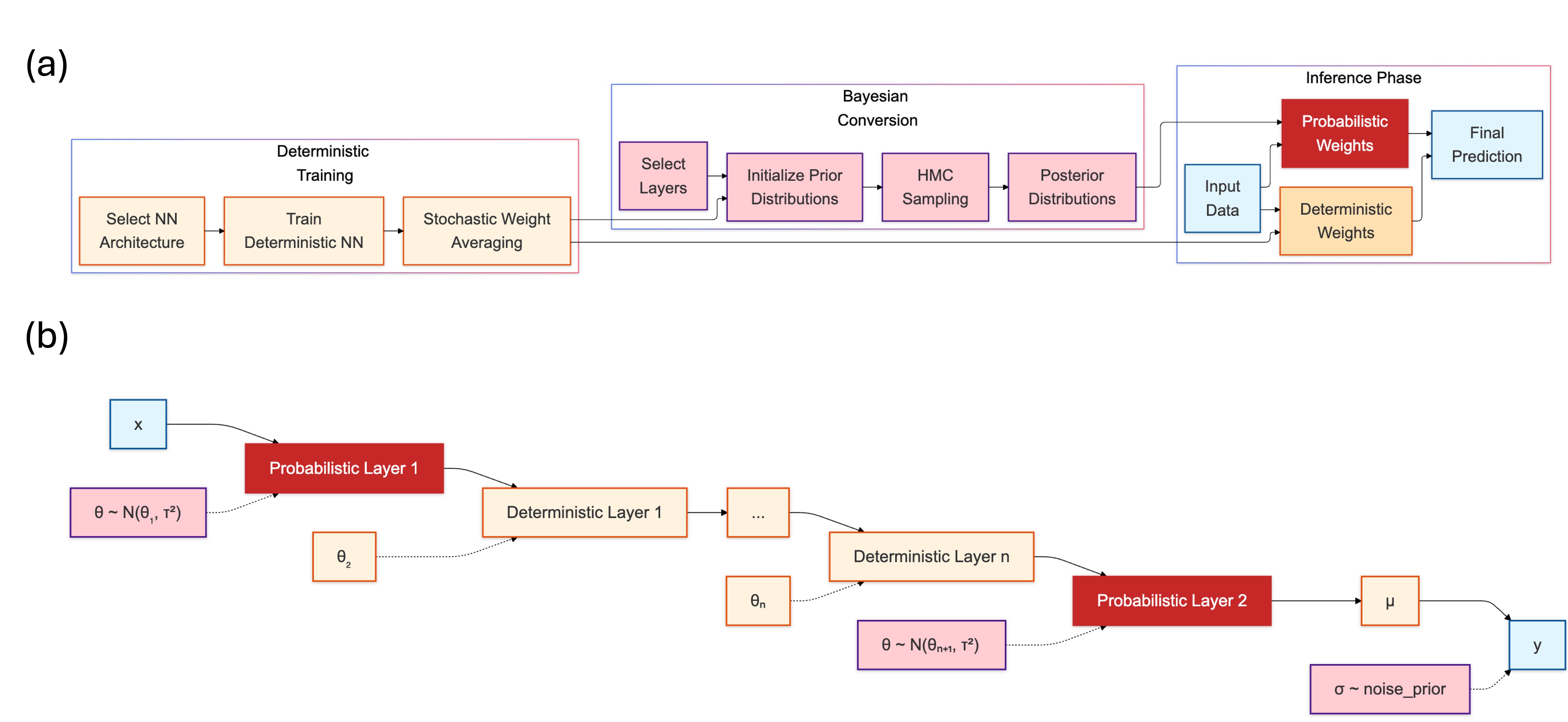}
    \caption{(a) Schematic illustration of Partially Bayesian Neural Network (PBNN) operation. First, we train a deterministic neural network, incorporating stochastic weight averaging to enhance robustness against noisy training objectives. Second, the probabilistic component is introduced by selecting a subset of layers and using the corresponding pre-trained weights to initialize prior distributions for this subset, while keeping all remaining weights frozen. HMC/NUTS sampling is then applied to derive posterior distributions for the selected subset. Finally, predictions are made by combining both the probabilistic and deterministic components. (b)  Schematic illustration of flow through a PBNN model alternating probabilistic and deterministic processing stages.
}
    \label{fig:PBNN_schema}
\end{figure}

The PBNNs are trained in two stages. First, it trains a deterministic neural network, incorporating stochastic weight averaging (SWA) \cite{izmailov2019averagingweightsleadswider} at the end of the training trajectory to enhance robustness against noisy training objectives. Second, the probabilistic component is introduced by selecting a subset of layers and using the corresponding pre-trained weights to initialize prior distributions for this subset, while keeping all remaining weights frozen. HMC/NUTS sampling is then applied to derive posterior distributions for the selected subset. Finally, predictions are made by combining both the probabilistic and deterministic components. See Algorithm \ref{alg:partial-bnn} and Figure \ref{fig:PBNN_schema} for more details. In certain scenarios, such as autonomous experiments, the entire training process needs to be performed in an end-to-end manner. In these cases, it is crucial to avoid overfitting in the deterministic component, as there will be no human oversight to evaluate its results before transitioning to the probabilistic part. To address this, we incorporate a MAP prior, modeled as a Gaussian penalty, into the loss function during deterministic training. All the PBNNs were implemented via a NeuroBayes package\footnote{\url{https://github.com/ziatdinovmax/NeuroBayes}} developed by the authors.

\begin{figure}[!htb]
    \centering
    \includegraphics[width=1.0\textwidth]{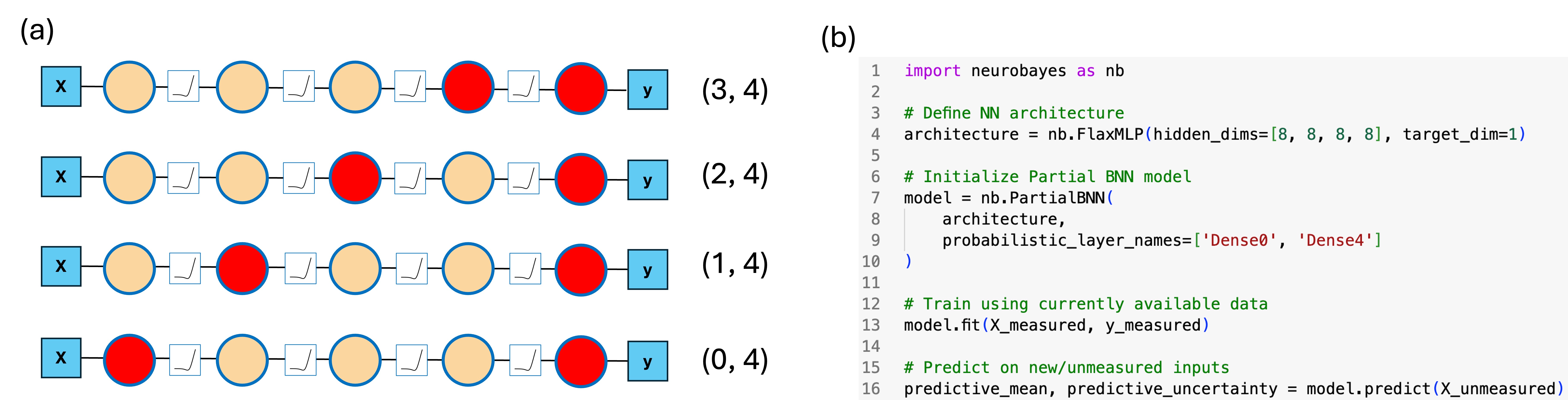}
    \caption{(a) Schematic representation of the partially Bayesian MLP employed in this study. The model consists of five layers: four utilize non-linear activation functions, such as the sigmoid linear unit, while the final (output) layer contains a single neuron without a non-linear activation, as is typical for regression tasks. Circles filled with red denote stochastic layers, while orange filled circles represent deterministic layers. Note that the single output neuron is always made probabilistic, as it often improves training stability. (b) Code snippet illustrating a single train-predict step with PBNN (0, 4).}
    \label{fig:PBNN_code}
\end{figure}

In this work, we have investigated PBNNs of multilayer perceptron (MLP) architecture consisting of five layers: four utilize non-linear activation functions, such as the sigmoid linear unit, while the final (output) layer contains a single neuron without a non-linear activation, as is typical for regression tasks. As there are multiple ways to select probabilistic layers for the PBNNs, we have evaluated the effects of setting different combinations of probabilistic layers as shown in Figure~ \ref{fig:PBNN_code}.

\begin{algorithm}[H]
\caption{Partially Bayesian Neural Network Training}
\label{alg:partial-bnn}
\begin{algorithmic}[0]
\Require
    \State Input data $X \in \mathbb{R}^{n \times d}$, targets $y \in \mathbb{R}^n$
    \State Deterministic neural network architecture $g_{\theta}$
    \State Set of probabilistic layers $\mathcal{L}$
    \State Optional: Custom SWA collection protocol $\psi$
    \State Optional: Custom prior width $\tau$ for probabilistic weights
    \State $\dagger$ Deterministic training hyperparameters follow typical deep learning practices
    \State $\ddagger$ Probabilistic training parameters follow standard Bayesian inference practices
\end{algorithmic}
\begin{algorithmic}[1]
\State Initialize network parameters $\theta$
\State Initialize empty weights collection $\mathcal{W} = \{\}$
\For{epoch $e = 1$ to $E$}
    \State $\eta_e, \text{collect} = \psi(e, E)$
    \State Update $\theta$ using SGD: $\theta \leftarrow \theta - \eta_e\nabla\mathcal{L}(\theta)$
    \If{$\text{collect}$}
        \State Add current weights to collection: $\mathcal{W} = \mathcal{W} \cup \{\theta\}$
    \EndIf
\EndFor
\State Compute averaged weights $\theta_{det} = \frac{1}{|\mathcal{W}|}\sum_{\theta \in \mathcal{W}} \theta$
\State // Run HMC/NUTS sampler for posterior inference
\For{each layer $l$ in network}
    \If{$l$ is probabilistic}
        \State Set prior $p(\theta_l) = \mathcal{N}(\theta_{det,l}, \tau)$
        \State Sample weights $\theta_l \sim p(\theta_l)$
    \Else
        \State Set weights $\theta_l = \theta_{det,l}$
    \EndIf
\EndFor
\State Calculate network output $\mu = g_{\theta}(X)$
\State Sample observation noise $\sigma \sim p(\sigma)$
\State Score observations $y \sim \mathcal{N}(\mu, \sigma^2)$
\State \Return Posterior samples of probabilistic weights and noise parameter
\end{algorithmic}
\end{algorithm}

\subsection{Active Learning}
In AL, the algorithm iteratively identifies points from a pool of unobserved data, within a pre-defined parameter space $\mathcal{X}_{\text{domain}} \subseteq \mathbb{R}^d$, that are expected to improve the model's performance in reaching some objective. Starting with an initial, usually small, training dataset $\mathcal{D} = \{ (x_i, y_i) \}_{i=1}^N$, an initial PBNN is trained and predictions are made on all $x^* \in \mathcal{X}_{\text{domain}}$. The predictions that maximize a suitably selected acquisition function are then selected for measurement via an experiment, simulation, or human labeling. For the sake of benchmarking, we have chosen an acquisition function that simply maximizes the predictive uncertainty, \emph{i.e.}, $x_{\text{next}} \gets \arg\max_{x^* \in \mathcal{X}_{\text{domain}} } U(x^*) $, and only select a single $x_{\text{next}}$ at each iteration. Note that here we naturally balance exploration between regions of model uncertainty and inherent complexity, as high aleatoric uncertainty often indicates areas requiring additional samples to better estimate noise distributions and capture underlying patterns. For further details regarding the AL algorithm, see Algorithm \ref{alg:active-learning}. Usually, this process is repeated until a desired goal is reached or an experimental budget is exhausted; here, we perform 200 exploration steps for all datasets. Lastly, we have selected initial training datasets by randomly sampling subsets of the total datasets containing 5\% of the total number of data points. While this procedure results in differently sized initial training datasets, the trends observed are consistent across all datasets and corresponding sizes.

\begin{algorithm}[H]
\caption{Active Learning}
\label{alg:active-learning}
\begin{algorithmic}[0]
\Require
    \State Parameter space \( \mathcal{X}_{\text{domain}} \subseteq \mathbb{R}^d \)
    \State Number of initial measurements \(N\)
    \State PBNN model architecture and parameters
    \State Stopping criterion
\end{algorithmic}
\begin{algorithmic}[1]
\State Conduct \(N\) random measurements to create initial dataset \( \mathcal{D} = \{ (x_i, y_i) \}_{i=1}^N \)
\State Train the PBNN on \( \mathcal{D} \) using Algorithm~\ref{alg:partial-bnn}
\Repeat
    \State Compute PBNN's predictive uncertainty \( U(x^*) \) for each \( x^* \in \mathcal{X}_{\text{domain}} \)
    \State \( x_{\text{next}} \gets \arg\max_{x^* \in \mathcal{X}_{\text{domain}} } U(x^*) \)
    \State Perform measurement at \( x_{\text{next}} \) to obtain \( y_{\text{next}} \)
    \State Update \( \mathcal{D} \) by adding \( (x_{\text{next}}, y_{\text{next}}) \)
    \State Re-train the PBNN on updated \( \mathcal{D} \) using Algorithm~\ref{alg:partial-bnn}
\Until{Stopping criterion is met}
\end{algorithmic}
\end{algorithm}

\subsubsection{Active Learning Metrics}
To assess the performance of active learning, we computed several key metrics after each active learning iteration. Our evaluation encompasses both prediction accuracy and uncertainty quantification. For each AL experiment, we have performed five runs with different random seeds to assess the robustness of our results. In each of the plots showing an AL metric as a function of AL step, a solid or dashed line denotes the mean of the metric across the five seeds, and the shaded region shows $\pm$1 standard deviation over the seeds, centered at the mean.

Prediction accuracy was evaluated using the standard root mean square error (RMSE):

\begin{equation} 
RMSE=\sqrt{\frac{\sum_{i}^{M}(y_{i}-{\mu}_{i})^2}{M}},
\label{eq:rmse}
\end{equation}
where $M$ is the size of the test set. 

To assess the quality of the predictive uncertainties, we used two metrics, the negative log predictive density (NLPD) and the confidence interval coverage probability, which we refer to as coverage from this point forward. NLPD  is given by the following equation:
\begin{equation}
    \text{NLPD} = -\frac{1}{M} \sum_{i=1}^M \left[ -\frac{1}{2} \log(2\pi U_i) - \frac{(y_i -{\mu}_{i})^2}{2 U_i} \right]
    \label{eq:nlpd}
\end{equation}
NLPD assesses how well a model's predictive distributions align with observed data. A lower NLPD indicates that the model assigns higher probability density to true outcomes while maintaining well-calibrated uncertainty estimates. This metric is valuable for evaluating probabilistic models as it penalizes both overconfident incorrect predictions and underconfident correct ones.

Coverage is given by
\begin{equation} \label{eq:coverage}
Coverage = \frac{1}{M}\sum_{i}^{M} \indicator_{y_{i}\in \text{CI}(x_{i})},
\end{equation}
where $\text{CI}(x_{i})$ is the confidence interval of test point $x_{i}$. Coverage measures the empirical reliability of a model's uncertainty estimates by calculating the proportion of true values that fall within the predicted confidence intervals, \emph{i.e.}, how often the true $y$ lies in the ML prediction inverval given by the predictive mean $\mu_{pred}$ and uncertainty $U_{pred}$. \cite{Coverage2021Beam, Sluijterman_Coverage} In this work, all coverage values are computed for 95\% confidence intervals. Coverages below 95\% indicate overconfident predictions (intervals too narrow) and coverages above 95\% indicate more conservative confidence intervals (intervals too wide), with a coverage value of 95\% being ideal. In practice, given the uneven costs of errors, models that produce a slightly conservative coverage are typically favored over those yielding overconfident assessments.

\subsubsection{Datasets}
To assess the performance of PBNNs for AL on a variety of diverse datasets, we have selected two molecular and two materials datasets for benchmarking, and one molecular and one materials dataset containing both simulation and experimental data to investigate transfer learning (TL) from computed to experimental properties. Details, such as the dataset sizes and relevant references, regarding these datasets are provided in Tables~\ref{tab:Datasets1} and~\ref{tab:Datasets2}. The FreeSolv, ESOL, and Steel fatigue (NIMS) datasets were used as published, while the Conductivity (HTEM) and Bandgap datasets are subsets of the published databases. Specifically, the Conductivity (HTEM) dataset utilized here is restricted to oxides containing Ni, Co, and Zn which have electrical conductivity values, and the Bandgap dataset is a random sample of 1000 non-metals from the intersection of the Materials Project bandgap dataset and the Matbench experimental bandgap dataset. We also used a noisy version of FreeSolv (Noisy-FreeSolv) for TL where experimental target values were corrupted by a zero-centered Gaussian noise with a standard deviation of one.

As far as the input features are concerned, we used standard RDKit \cite{rdkit2022rdkit} physicochemical descriptors for the molecular datasets. For the steel fatigue dataset, the input features were chemical compositions, upstream processing details, and heat treatment conditions. For the electrical conductivity data, the input features were formed from oxide concentrations, deposition conditions, and processing parameters, such as power settings and gas flow rate. The input features for the Bandgap dataset were derived using the Magpie featurizer, which computes statistical descriptors from elemental properties and composition fractions \cite{magpie2016}.

\begin{table}[h]
    \centering
    \caption{Datasets for Active Learning}
    \begin{tabular}{c|c|c|c|c}
    \hline
    \textbf{Name} & \textbf{Target property} & \textbf{$N_{features}$} & \textbf{$N_{samples}$} & \textbf{Reference} \\ \hline
    FreeSolv & Hydration free energy & 9 & 642 & \cite{Mobley_FreeSolv} \\ \hline
    ESOL & Aqueous solubility & 9 & 1128 & \cite{Delaney_ESOL} \\ \hline
    Steel fatigue (NIMS) & Fatigue strength & 25 & 437 & \cite{Agrawal_Fatigue} \\ \hline 
    Conductivity (HTEM) & Electrical conductivity & 12 & 1184 & \cite{Zakutayev_HTEM} \\ \hline
    \end{tabular}
    \label{tab:Datasets1}
\end{table}

\begin{table}[h]
    \centering
    \caption{Datasets for Transfer Learning}
    \begin{tabular}{c|c|c|c|c}
    \hline
    \textbf{Name} & \textbf{Target property} & \textbf{$N_{features}$} & \textbf{$N_{samples}$} & \textbf{Reference} \\ \hline
    Noisy-FreeSolv & Hydration free energy & 9 & 642 & \cite{Mobley_FreeSolv} \\ \hline
    Bandgap & Bandgap energy & 132 & 1000 & \cite{Jain_MP, Zhuo_Expt_Gap} \\ \hline
    \end{tabular}
    \label{tab:Datasets2}
\end{table}

\section{Results and Discussion}

\subsection{Active learning on toy dataset}
Before assessing the effectiveness of PBNNs for AL on the materials and molecular datasets, we first analyze their effectiveness on a toy dataset. In particular, we have generated non-stationary data with abrupt changes in frequency and amplitude, a use case where Full BNNs consistently outperform GPs \cite{ziatdinov2024fullybayesian}. We denote PBNN configurations as PBNN (\textit{i}, 4), where \textit{i} indicates which hidden layer is probabilistic (counting from 0), and 4 denotes the output layer that is always treated as probabilistic. For example, PBNN (0, 4) has probabilistic first hidden and output layers, while PBNN (3, 4) has probabilistic last hidden and output layers. To ensure fair comparison, all hidden layers have equal width. The output layer consists of a single neuron, so making it probabilistic adds minimal computational overhead while helping with training stability. 
\begin{figure}[htbp!]
    \centering
    \includegraphics[width=0.8\textwidth]{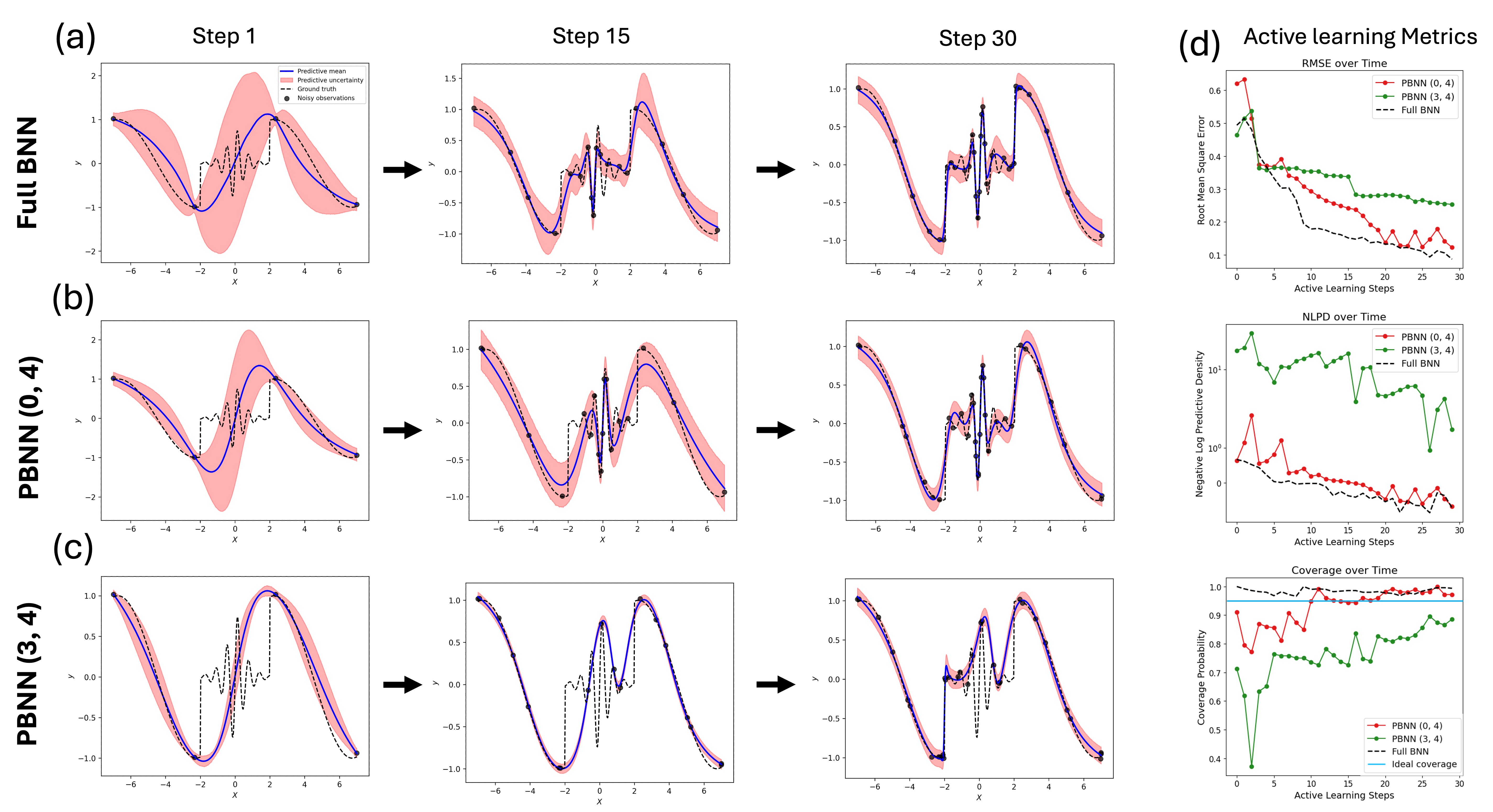}
    \caption{Active learning results for 1D toy dataset with non-stationary features. Evolution of predictions and uncertainty estimates across active learning steps for (a) Full BNN model, (b) model with probabilistic first hidden and output layers, PBNN(0,4), and (c) model with probabilistic last hidden and output layers, PBNN(3,4). Blue lines show predictive mean, pink shading represents uncertainty estimates, and black dashed lines indicate ground truth. Black dots mark noisy observations. Panel (d) compares performance metrics (RMSE, NLPD, and coverage probability) as a function of active learning exploration steps.}
    \label{fig:PBNN_toydata}
\end{figure}
Figure \ref{fig:PBNN_toydata} shows the evolution of predictions and uncertainty estimates across active learning steps for Full BNN (a), PBNN (0,4) (b), and PBNN (3,4) (c). PBNN (0,4) exhibits behavior remarkably similar to Full BNN, both in terms of predictive mean and uncertainty estimates (shown as pink shading), while requiring fewer probabilistic layers. In contrast, PBNN (3,4) struggles to provide reliable uncertainty estimates, particularly in regions with strong oscillatory behavior. This is reflected in the performance metrics (d), where PBNN (0,4) closely tracks Full BNN's performance while PBNN (3,4) shows consistently higher RMSE and NLPD values, along with coverage probability further from the ideal value of 0.95.

\subsection{Active learning on molecular datasets}
\begin{figure}[htbp!]
    \centering
    \includegraphics[width=1.0\textwidth]{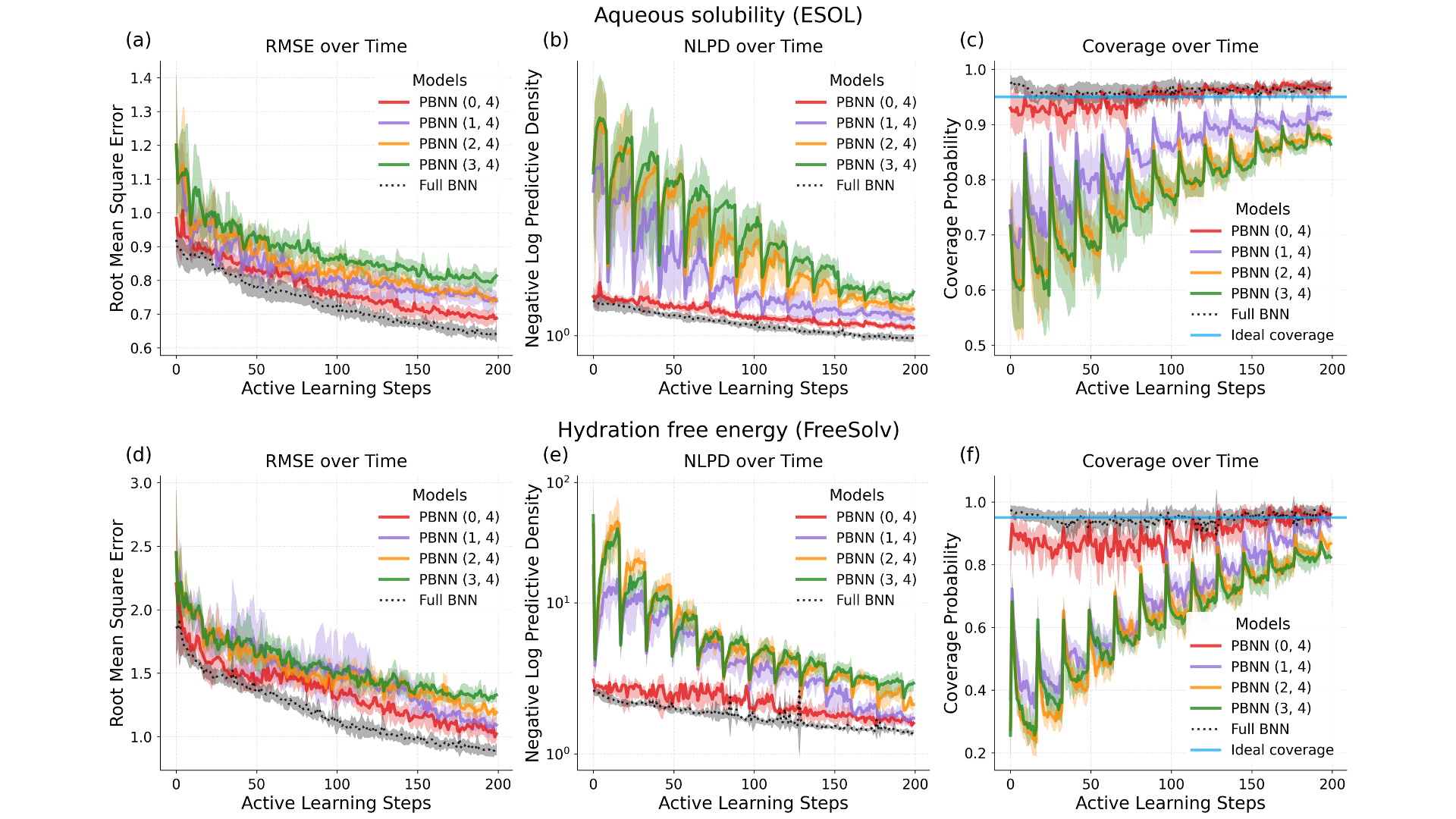}
    \caption{ Comparison of Partially Bayesian Neural Networks (PBNNs) and fully Bayesian neural network (Full BNN) on molecular property prediction tasks. (a) Aqueous solubility prediction (ESOL database) and (b) hydration free energy prediction (FreeSolv database). Each PBNN configuration PBNN (i, 4) has two probabilistic layers: one at position i (counting from 0) and one at the output. Shaded areas represent a standard deviation across five different random seeds.}
    \label{fig:PBNN_molecules}
\end{figure}

We now investigate the effectiveness of different PBNNs for AL on the standard molecular benchmark datasets. Figures \ref{fig:PBNN_molecules}(a) and \ref{fig:PBNN_molecules}(b) show RMSE, NLPD, and coverage probability as a function of AL exploration step for ESOL and FreeSolv, respectively. We see that the accuracy and quality of the uncertainties improve with AL for all PBNNs, as demonstrated by \emph{i}) decreasing RMSE and NLPD and \emph{ii}) coverage approaching 0.95 for all models. Across all metrics for both datasets, making earlier layers probabilistic proves more effective, with PBNN(0,4) approaching the accuracy of a Full BNN. Furthermore, PBNN(0,4) exhibits a relatively stable decrease in NLPD and coverage approaching 0.95 throughout the AL process, similar to Full BNN. In contrast, configurations where the probabilistic layer is moved away from the first hidden layer, PBNN(1,4), (2,4), and (3,4), show strong oscillatory behavior in NLPD and coverage metrics, suggesting that uncertainty propagation becomes unstable when probabilistic layers are placed in later hidden layers. This shows that, at least within the standard MLP architecture employed here, capturing uncertainty in the first feature transformation layer, combined with a probabilistic output layer, is more effective, both in terms of performance and reliability. In addition, it decreased the overall computational time by nearly a factor of four. Notably, with only a fraction of points explored, AL with PBNN achieves accuracy either comparable to (ESOL) or better than (FreeSolv) that obtained using standard 80:20 or 90:10 train-test splits with standard deterministic ML models \cite{wu2018moleculenetbenchmarkmolecularmachine}.
\subsection{Active learning on materials datasets}
\begin{figure}[htbp!]
    \centering
    \includegraphics[width=1.0\textwidth]{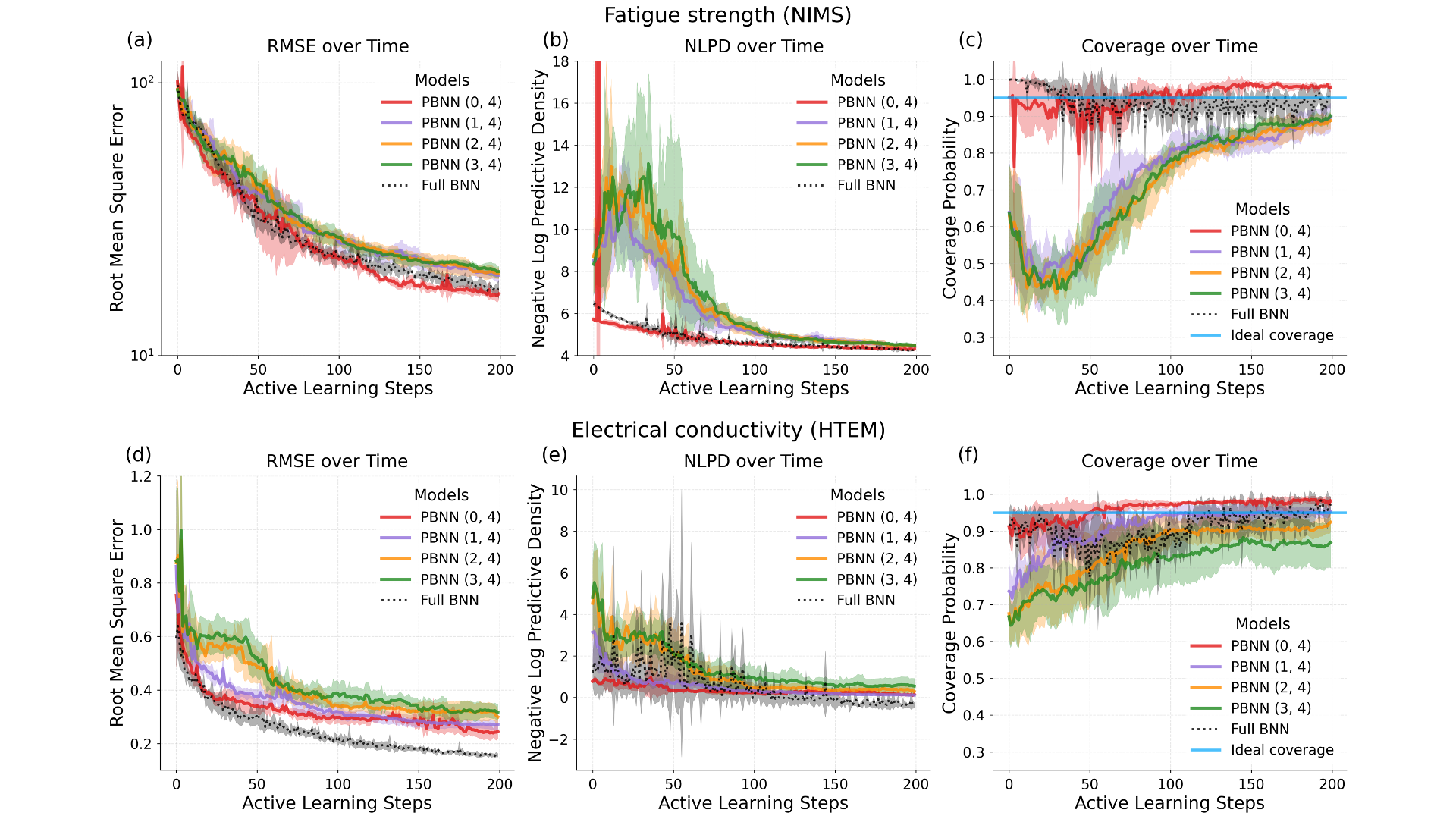}
    \caption{ Comparison of Partially Bayesian Neural Networks (PBNNs) and fully Bayesian neural network (Full BNN) on materials property prediction tasks. (a) Fatigue strength prediction (NIMS database) and (b) electrical conductivity prediction (HTEM database). Each PBNN configuration PBNN (i, 4) has two probabilistic layers: one at position i (counting from 0) and one at the output. Shaded areas represent a standard deviation across five different random seeds.}
    \label{fig:PBNN_materials}
\end{figure}

Next, we follow a similar analysis for the two materials datasets, Steel fatigue (NIMS) and Conductivity (HTEM), as shown in Figure \ref{fig:PBNN_materials}. We observe overall similar trends to the molecular datasets (decreasing RMSE and NLPD and coverage approaching 0.95), although we see a much stronger difference between the different PBNNs in the uncertainty metrics, with smaller difference in RMSE across different selections of probabilistic layers. We also do not observe the clean monotonic trends that we observed with the molecular datasets for NLPD and Coverage on the Steel fatigue (NIMS) dataset. This could be due to a variety of factors, but we suspect that this is largely due to differences in the types of input features. While the molecular datasets utilized SMILES-derived descriptors as their input features, the materials datasets contained experimental parameters as their input features, which may not be as predictive of the target properties as the structural SMILES-based descriptors. There could also be a difference in experimental noise between the molecular and materials datasets, as it is well known that values of the materials target properties, fatigue strength and electrical conductivity, are sensitive to experimental variations in their measurement, whereas measurements of hydration free energy and aqueous solubility are relatively standardized. 

Despite these domain-specific variations, the results across both molecular and materials domains support the emerging general principle that making the first hidden and the output layers probabilistic is more effective than doing so for intermediate or final layers.  We would also like to emphasize that we used the same MLP architecture and training parameters (SGD learning rate and iterations for the deterministic component, warmup steps and samples for NUTS in the probabilistic component) across all four datasets. This demonstrates that PBNNs can be relatively robust to hyperparameter selection, a valuable characteristic for practical applications as it minimizes the need for extensive dataset-specific tuning.

\subsection{Convergence diagnostics}
We next discuss convergence diagnostics for PBNN models during active learning. A popular choice for convergence diagnostics in Bayesian inference is the Gelman-Rubin statistic (‘R-hat’), which provides a measure of convergence for each model parameter \cite{GelmanRubin1992}. However, for Bayesian neural networks, where the parameter space is high-dimensional, examining individual parameter convergence becomes impractical. Instead, we analyzed the distribution of R-hat values across all parameters and found that for the majority of weights (95--99\%, depending on dataset), these values lie within acceptable ranges between 1.0 and 1.1 \cite{BrooksGelman1998}. While layer-wise or module-wise convergence analysis is also possible for complex architectures, we opted for global parameter statistics due to the relatively simple network structure in this study. See Appendix 1 for more details.

We note that in active learning-based autonomous science tasks, reliable convergence diagnostics play an important role in ensuring the autonomous system performance. The R-hat statistic can therefore serve as an automated quality check, triggering specific actions when convergence issues are detected: for example, if a high proportion (> 10\%) of parameters display R-hat values outside the acceptable range, the system can employ various convergence improvement heuristics. These include increasing the number of warm-up states, trying different parameter initialization schemes, or adjusting prior distributions. If issues persist after these interventions, the system can flag the experiment for human review, ensuring reliability of the autonomous decision-making process.

\subsection{Transfer learning}
Transfer learning (TL) is a machine learning method commonly used to improve model performance and/or accelerate training by leveraging knowledge from a related task. TL is particularly valuable when data is limited and difficult to acquire, as is often the case in experimental materials science and chemistry. For deterministic NNs, TL is performed by initializing the network parameters with those of a pre-trained network. Most often the target NN's parameters are still optimized for the task at-hand via backpropagation, which is referred to as fine-tuning.

In the context of BNNs, transfer learning can be done through a selection of prior distributions over weights. First, let us revisit how the initial prior distributions are commonly selected. Ideally, the goal of priors in the Bayesian framework is to have a principled way to incorporate domain knowledge. In practice, however, we simply set priors of BNNs to zero-centered normal distributions. This approach provides good regularization and meaningful uncertainty estimates in predictions, but it doesn't incorporate any actual prior domain knowledge. We argue that we can use the weights of a deterministic model trained in a computational ("digital twin") space to initialize these prior distributions by setting their means to the corresponding pre-trained weights, thereby transferring domain knowledge to a (P)BNN operating in the real world. We can choose to do it for the entire model or only for some parts (layers) of it. We can also specify a "degree of trust" in the theory by selecting appropriate standard deviations for these distributions: wider distributions indicate less confidence in the computational model, while narrower ones encode stronger belief in the underlying theory.

\begin{figure}[htbp!]
    \centering
    \includegraphics[width=0.6\textwidth]{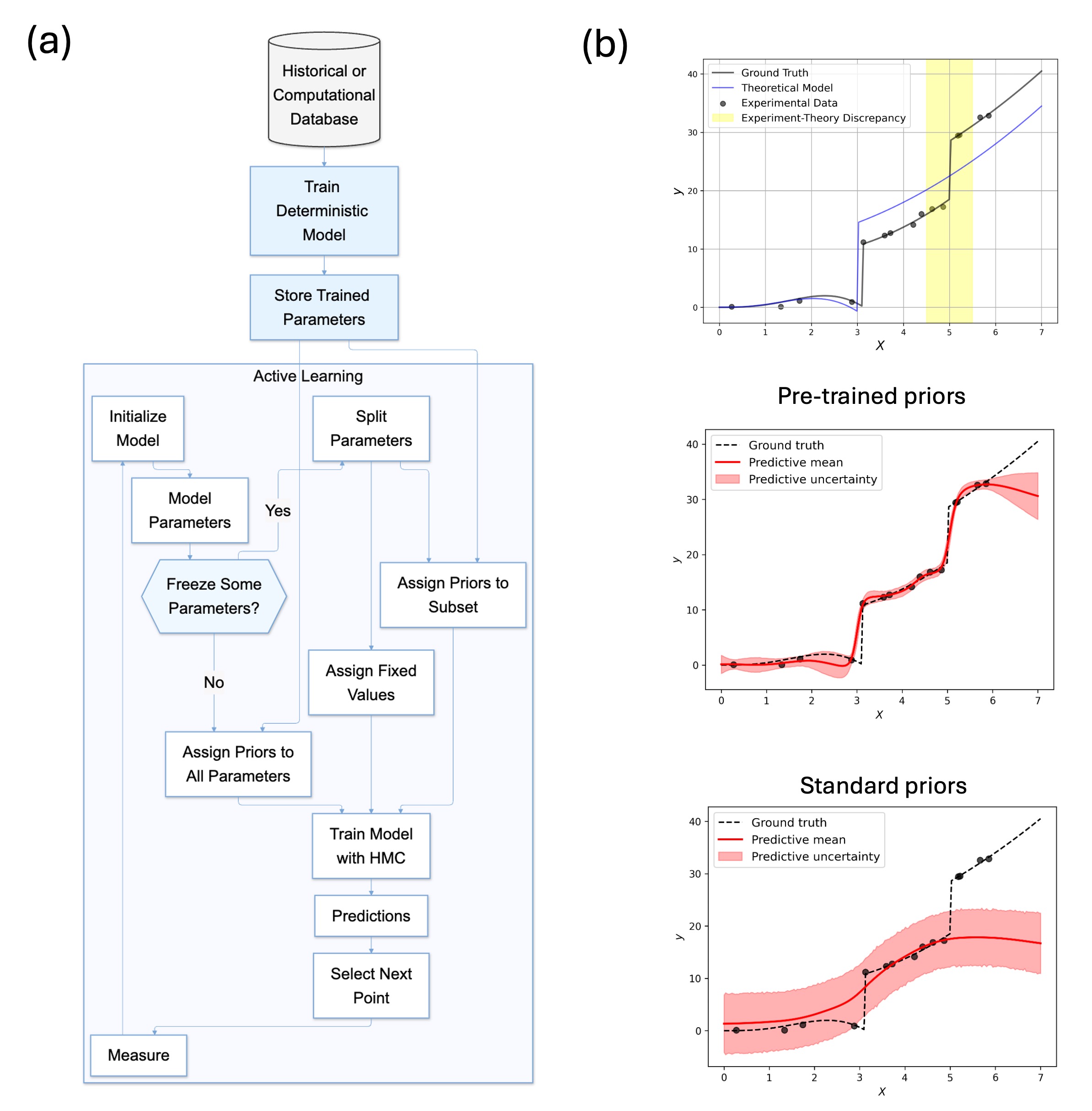}
    \caption{(a) Schematic workflow of transfer learning in active learning: a deterministic model is first trained on historical or computational data, and its parameters are used to initialize and inform priors in the subsequent active learning process. (b) Example predictions comparing BNNs with pre-trained and standard priors.  
}
    \label{fig:pretrained_priors_algo}
\end{figure}

Here, we examine how this simulation-to-experiment transfer learning affects AL with (P)BNNs. The process involves first training a deterministic NN on simulation data, then using its weights to inform the (P)BNN surrogate model that guides active learning on experimental data. A schematic showing this joint TL-AL process is shown in Figure \ref{fig:pretrained_priors_algo} (a). 
We first study the effectiveness of TL via PBNNs using 1D toy data, where we have generated "theoretical" and "experimental" datasets, emulating phase transitions, with the latter introducing an abrupt change in behavior not captured by the theoretical model (Figure \ref{fig:pretrained_priors_algo} (b)). This is a rather common scenario in scientific modeling where theoretical models reflect the overall trend but fail to capture certain experimental phenomena. Such discrepancies are frequently encountered when theoretical models rely on simplifying assumptions and fail to account for complex physical mechanisms. The middle and bottom panels in Figure \ref{fig:pretrained_priors_algo} (b) compare two approaches to addressing this challenge using full BNNs. Using pre-trained priors, informed by the theoretical model, allows the BNN to maintain good predictions in regions where the theory works well while adapting to experimental evidence where it doesn't. In contrast, standard uninformative priors fail to capture a second phase transition and lead to overly conservative uncertainty estimates.
\begin{figure}[htbp!]
    \centering
    \includegraphics[width=1.0\textwidth]{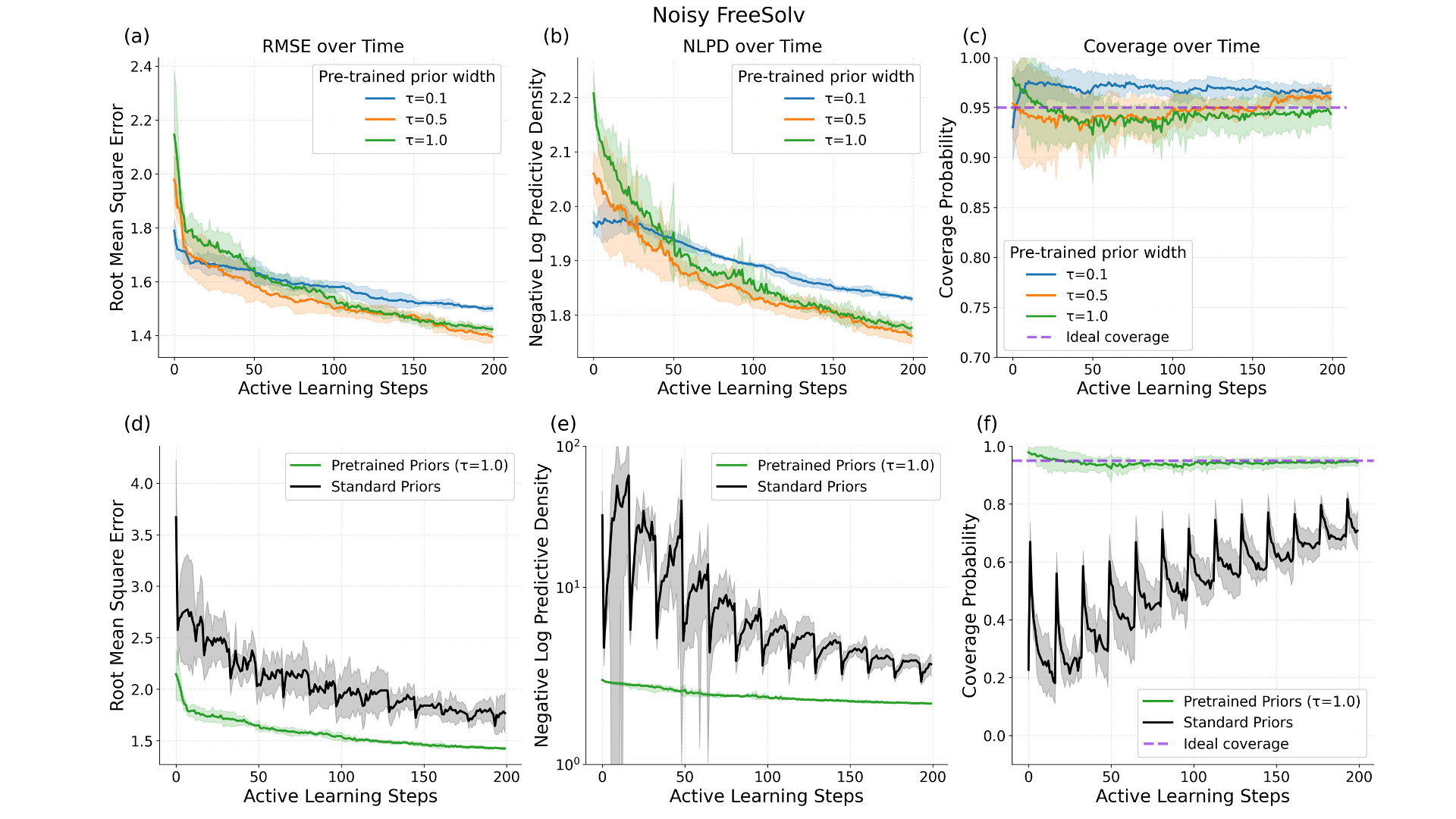}
    \caption{Transfer learning with pre-trained PBNNs applied to noisy FreeSolv dataset. (a ) RMSE, NLPD, and coverage probability for different prior widths ($\tau$). (b) Comparing the performance of pre-trained priors ($\tau=1.0$) against standard priors. Shaded areas represent a standard deviation across five different random seeds.}
    \label{fig:PBNN_TL_FreeSolv}
\end{figure}

We now move to the molecular and materials datasets. We start with the Noisy-FreeSolv dataset. Here the deterministic neural network is pre-trained on computational data from molecular dynamics simulations, whereas experimental data is augmented with synthetic noise to create a more challenging test case for our models. For this study, we made the last two hidden layers and the output layer probabilistic, with priors initialized at values of weights from the corresponding pre-trained deterministic neural network. Figure \ref{fig:PBNN_TL_FreeSolv} shows the performance of PBNN with theory-informed priors for different prior widths ($\tau$). While all prior widths demonstrate good performance, wider priors ($\tau=0.5, 1.0$) outperform narrower priors $\tau=0.1$ across all metrics. This can be explained by the fact that with small $\tau$ values, the prior (informed by the theoretical model) dominates the likelihood in shaping the posterior, failing to account for discrepancy between experiment and theory, whereas with larger values, the likelihood (based on experimental data) starts exerting greater influence. Overall, an ideal value would balance leveraging theoretical knowledge and adapting to experimental observations for a given set of theoretical and experimental data. Comparing pre-trained and standard priors at $\tau=1.0$, we observe that theory-informed priors lead to substantially better performance across all metrics. The improvement is particularly pronounced in NLPD and coverage, where standard priors show high uncertainty and unstable behavior throughout the active learning process.
\begin{figure}[htbp!]
    \centering
    \includegraphics[width=1.0\textwidth]{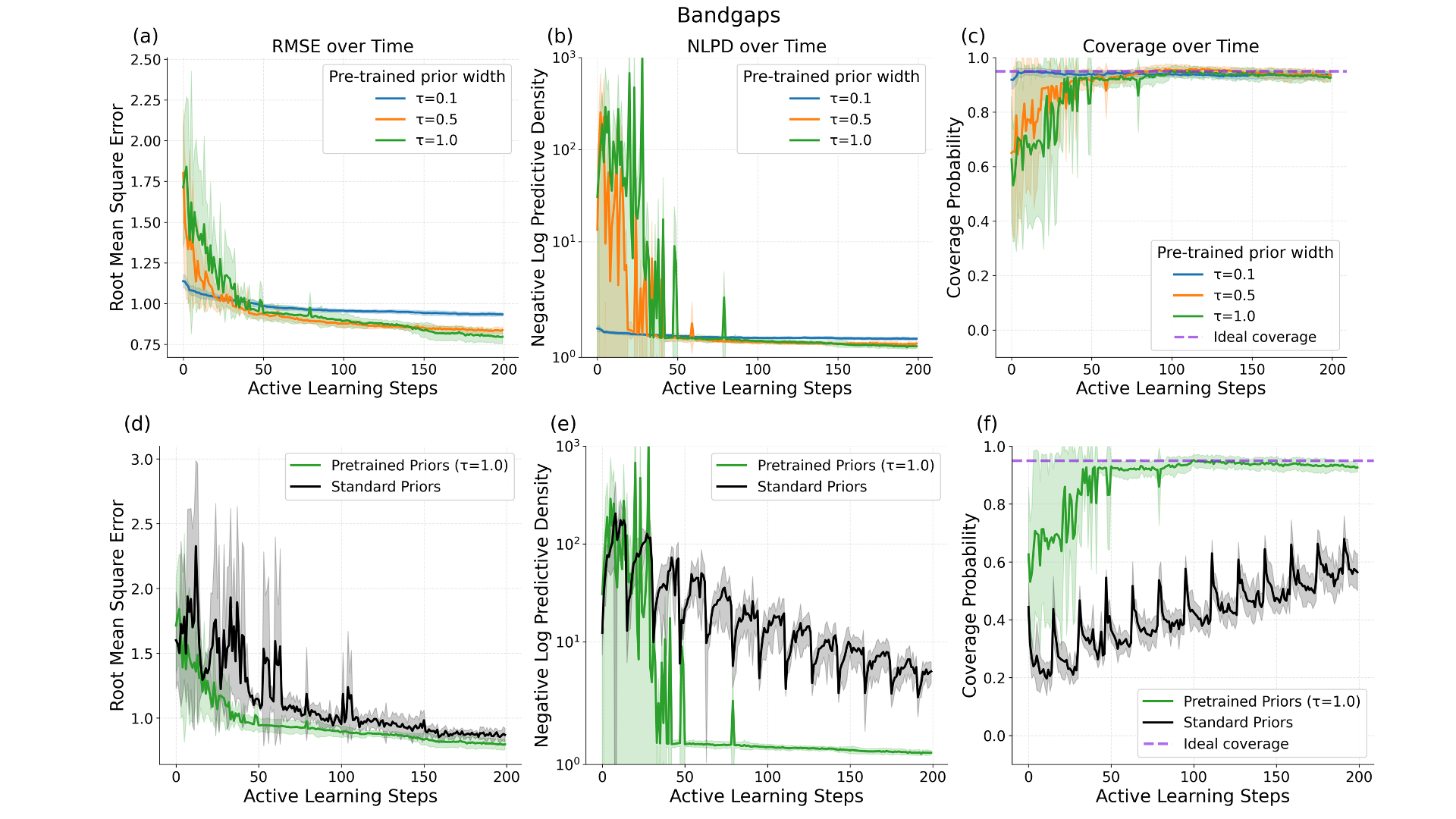}
    \caption{Transfer learning with pre-trained PBNNs applied to Bandgaps dataset. (a ) RMSE, NLPD, and coverage probability for different prior widths ($\tau$). (b) Comparing the performance of pre-trained priors ($\tau=1.0$) against standard priors. Shaded areas represent a standard deviation across five different random seeds.}
    \label{fig:PBNN_TL_Bandgap}
\end{figure}
Finally, we analyze bandgaps of non-metals, where priors are pre-trained on density functional theory (DFT) calculations. Similar to the results observed for Noisy FreeSolv, the results shown in Figure~\ref{fig:PBNN_TL_Bandgap} demonstrate that among different prior widths, there is a clear trade-off: the tight prior ($\tau=0.1$) shows stable but limited improvement, suggesting it constrains the model too closely to DFT predictions, while wider priors ($\tau=0.5$ and $\tau=1.0$) show initial oscillations but ultimately achieve better RMSE through greater adaptation to experimental data. This suggests that one can in principle apply dynamic adjustment: impose a strong belief in the theoretical model initially, and then, as more data becomes available, gradually relax it, allowing the data to speak for itself. Comparing pre-trained and standard priors at $\tau=1.0$, we observe similar trends to the FreeSolv dataset. The advantage of pre-trained priors is particularly pronounced in the early stages of active learning, where in the first 50 steps they achieve significantly lower RMSE and better calibrated uncertainties compared to standard priors, indicating more efficient use of limited experimental data. While both approaches eventually converge to similar RMSE values, the benefits of pre-trained priors persist in uncertainty quantification throughout the entire process, maintaining substantially better coverage probability.

\section{Conclusion}
In this work, we explored the capabilities of partially Bayesian neural networks (PBNNs) in active learning tasks. Within the MLP architectures deployed here, we found that the choice of which layers are made probabilistic significantly impacts performance, with early layers providing better and more stable uncertainty estimates - a finding that held consistently across studied molecular and materials datasets. Notably, PBNNs with probabilistic first  layer achieved performance comparable to fully Bayesian networks while requiring substantially fewer computational resources.

We further enhanced PBNN performance through transfer learning by initializing priors using theoretical models, which proved particularly beneficial in the early stages of active learning. Our analysis revealed an important trade-off in prior width selection: tight priors ensure stability but may constrain the model too closely to theoretical predictions, while wider priors enable better adaptation to experimental data. Across both studied systems, theory-informed priors led to better calibrated uncertainties and more efficient data utilization.

Overall, this work demonstrates the feasibility of PBNNs for materials science and chemistry, particularly in the context of AL for limited, complex datasets. In the future, we plan to  explore the effectiveness of PBNNs across a wider range of architectures, including Graph Neural Networks and Transformers. There may also be interesting connections between our findings about which layers to make probabilistic and emerging research on neural network interpretability, particularly studies that identify specific neurons responsible for distinct behaviors in large language models. Just as certain neurons in LLMs have been found to encode specific linguistic features or semantic concepts, understanding which neurons are most critical for uncertainty quantification could inform more targeted and efficient PBNN architectures.

\section{Code and Data Availability}
Code and data supporting the paper's findings, together with additional implementation details, are available at 
\url{https://github.com/ziatdinovmax/NeuroBayes/tree/paper/active_learning_scripts}

\section*{Acknowledgments}
This work was supported by the Laboratory Directed Research and Development Program at Pacific Northwest National Laboratory, a multiprogram national laboratory operated by Battelle for the U.S. Department of Energy.

\printbibliography  






\clearpage
\appendix
\renewcommand{\thesection}{}  
\section{Appendix 1}
\renewcommand{\thefigure}{A\arabic{figure}}
\setcounter{figure}{0}  

\begin{figure}[htbp]
    \centering
    \includegraphics[width=1.0\textwidth]{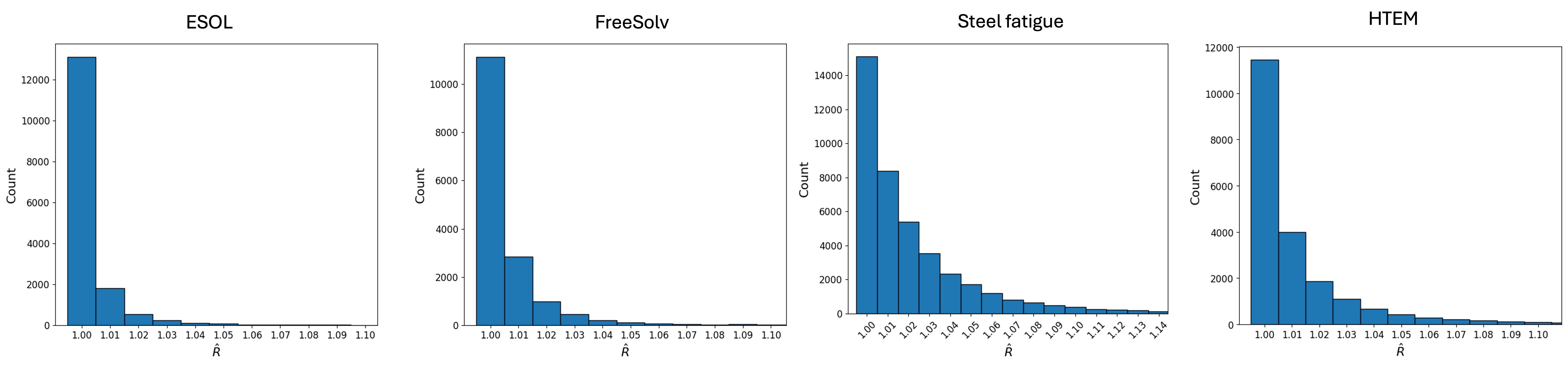}
    \caption{Gelman-Rubin ‘R-hat’ values over all active learning steps for ESOL, FreeSolv, Steel fatigue, and HTEM datasets.}
    \label{fig:A1}
\end{figure}

Figure~\ref{fig:A1} shows the distribution of R-hat values across PBNN (0, 4) parameters aggregated over all active learning steps for four different case studies: ESOL, FreeSolv, Steel fatigue, and HTEM datasets. All cases demonstrate good convergence characteristics, with the majority of parameters having R-hat values close to 1.0. The distributions exhibit a right-skewed pattern, which is expected in MCMC convergence diagnostics. There are, however, variations between datasets - particularly, the Steel fatigue case shows a wider spread of R-hat values, which correlates with more volatile NLPD values and slower Coverage convergence in early active learning steps. Nevertheless, most of the weights and biases fall within the the range 1.0 < R-hat < 1.1, which is traditonally considered to indicate good convergence. 

\end{document}